\documentclass[12pt]{article}
\usepackage{amsmath}
\usepackage{graphicx}

\begin{document}

%%%%%%%%%%%%%%%%%%%%%%%%%%%%%%%%%%%%%%%%%%%%%%%%%%%%%%%%%%%%%%%%%%%%
% basic data for the eprint:
%%%%%%%%%%%%%%%%%%%%%%%%%%%%%%%%%%%%%%%%%%%%%%%%%%%%%%%%%%%%%%%%%%%%

\textwidth=6.0in  \textheight=8.25in

%%  Adjust these for your printer:
\leftmargin=-0.3in   \topmargin=-0.20in

%% preprint number data:
\newcommand\pubnumber{Nikhef-2010-044}
\newcommand\pubdate{\today}

%%  address and funding acknowledgement data:
\def\support{\footnote{Work supported by FOM}}

%%%%%%%%%%%%%%%%%%%%%%%%%%%%%%%%%%%%%%%%%%%%%%%%%%%%%%%%%%%%%%%%%%%%%%%%%%%%
%   document style macros
%%%%%%%%%%%%%%%%%%%%%%%%%%%%%%%%%%%%%%%%%%%%%%%%%%%%%%%%%%%%%%%%%%%%%%%%%%%%
\def\Title#1{\begin{center} {\Large #1 } \end{center}}
\def\Author#1{\begin{center}{ \sc #1} \end{center}}
\def\Address#1{\begin{center}{ \it #1} \end{center}}
\def\andauth{\begin{center}{and} \end{center}}
\def\submit#1{\begin{center}Submitted to {\sl #1} \end{center}}
\newcommand\pubblock{\rightline{\begin{tabular}{l} \pubnumber\\
         \pubdate  \end{tabular}}}
\newenvironment{Abstract}{\begin{quotation}  }{\end{quotation}}
\newenvironment{Presented}{\begin{quotation} \begin{center} 
             PRESENTED AT\end{center}\bigskip 
      \begin{center}\begin{large}}{\end{large}\end{center} \end{quotation}}
\def\Acknowledgements{\bigskip  \bigskip \begin{center} \begin{large}
             \bf ACKNOWLEDGEMENTS \end{large}\end{center}}
%%%%%%%%%%%%%%%%%%%%%%%%%%%%%%%%%%%%%%%%%%%%%%%%%%%%%%%%%%%%%%%%%%%%%%%%%%%%
%   Title page
%%%%%%%%%%%%%%%%%%%%%%%%%%%%%%%%%%%%%%%%%%%%%%%%%%%%%%%%%%%%%%%%%%%%%%%%%%%%

\thispagestyle{empty}

\begin{flushright}
Nikhef-2010-044\\
\end{flushright}

\vspace{0.0truecm}
\begin{center}
\boldmath
\large\bf Extracting $\gamma$ and Penguin Parameters\\
from $B_s^0\to J/\psi K_{\rm S}$
\unboldmath
\end{center}

\vspace{0.3truecm}
\begin{center}
Kristof De Bruyn, Robert Fleischer  \,and\, Patrick Koppenburg\\[0.1cm]
{\sl Nikhef, Science Park 105, 
NL-1098 XG Amsterdam, The Netherlands}
\end{center}

\vspace{0.3truecm}

\begin{center}
{\bf Abstract}
\end{center}

{\small
\vspace{0.2cm}\noindent
The $B_s^0\to J/\psi K_{\rm S}$ decay has recently been observed by the CDF 
collaboration and will be of interest for the LHCb experiment. It is the $U$-spin partner
of the ``golden" $B_d^0\to J/\psi K_{\rm S}$ channel and offers a determination of
the angle $\gamma$ of the unitarity triangle. Moreover, it allows us to control the
hadronic penguin effects in the extraction of the $B^0_d$--$\bar B^0_d$ mixing phase
$\phi_d$ through a measurement of CP violation in $B_d^0\to J/\psi K_{\rm S}$. We 
discuss the picture emerging for these measurements from an LHCb feasibility study. While
LHCb will be able to extract $\gamma$ from the CP violation in $B^0_d\to J/\psi K_{\rm S}$, 
the main application of this channel will be the determination of hadronic penguin
parameters. Such an analysis is actually needed in order to fully exploit LHCb's impressive
experimental precision for the determination of $\phi_d$ from $B_d^0\to J/\psi K_{\rm S}$.
We give also the target regions for the effective lifetime of  $B_s^0\to J/\psi K_{\rm S}$ 
and its CP-violating observables for improved measurements by CDF and the 2011  
data taking at LHCb.
}

\vspace{0.7truecm}

\begin{center}
{\sl Talk at the 6th International Workshop\\ on the CKM Unitarity Triangle
(CKM2010)\\
Warwick, United Kingdom, 6--10 September 2010\\
To appear in the Proceedings}
\end{center}

\vfill
\noindent
December 2010

\newpage
\thispagestyle{empty}
\vbox{}
\newpage
 
\setcounter{page}{1}

%%%%%%%%%%%%%%%%%%%%%%%%%%%%%%%%%%%%%%%%%%%%%%%%%%%%%%%%%%%%%%%%%%%%%%%%%%%%
%   4 page text
%%%%%%%%%%%%%%%%%%%%%%%%%%%%%%%%%%%%%%%%%%%%%%%%%%%%%%%%%%%%%%%%%%%%%%%%%%%%

\section{Introduction}
%%%%%%%%%%%%%%%%%%%%%%%%%%%%%%%%%%%%%%%%%%%%%%%%%%%%%%%%%%%%%%%%%%%%%%%%%%%%
With LHCb taking data now, we have just entered a new era of precision flavour physics. 
This will allow us to implement new strategies for the extraction of the angle $\gamma$ of the 
unitarity triangle (UT). One such method is offered by the $B_s^0\to J/\psi K_{\rm S}$ mode
\cite{RF-BspsiK}, which has recently been observed by the CDF collaboration \cite{CDF-obs}. 
As was also noted in Ref.~\cite{RF-BspsiK}, this channel allows us moreover to include the 
contributions from penguins to the determination of the $B^0_d$--$\bar B^0_d$ mixing phase 
from the  ``golden" $B_d^0\to J/\psi K_{\rm S}$ decay. Including these effects will become 
mandatory in order to match LHCb's impressive precision. To accomplish this task, the $U$-spin 
flavour symmetry of strong interactions is used to relate the hadronic parameters of the 
$B_s^0\to J/\psi K_{\rm S}$ and $B_d^0\to J/\psi K_{\rm S}$ modes to one another, as the 
decay topologies of both channels are related by interchanging all down and strange quarks.

Here we shall give a short summary of the prospects for implementing these measurements
at LHCb. For a detailed discussion, including the derivations of the relevant formulae, an 
overview of the used feasibility study and further numerical results,  the reader is referred 
to Ref.~\cite{Bs2JpsiKs}.

\boldmath
\section{Extraction of $\gamma$}\label{sec:gam}
\unboldmath
%%%%%%%%%%%%%%%%%%%%%%%%%%%%%%%%%%%%%%%%%%%%%%%%%%%%%%%%%%%%%%%%%%%%%%%%%%%%
In the Standard Model, the $B_s^0\to J/\psi K_{\rm S}$ decay amplitude can be written 
as follows:
 \begin{equation}\label{Bs-ampl}
A(B_s^0\to J/\psi\, K_{\rm S})\propto\left[1-a e^{i\theta}e^{i\gamma}\right],
\end{equation}
where the CP-conserving parameters $a$ and $\theta$ characterise the penguin effects
with respect to the colour-allowed tree-diagram-like contribution, which plays the dominant
role. In order to determine the UT angle $\gamma$, as well as $a$ and $\theta$, the direct 
CP violation, $C$, the mixing-induced CP violation, $S$, and the observable $H\propto\text{BR}(B_s^0\to J/\psi K_{\rm S})/\text{BR}(B_d^0\to J/\psi K_{\rm S})$ have to be measured, which 
is the focus of our LHCb feasibility study. 

For the penguin parameters, we use the results obtained
in Ref.~\cite{FJFM} from an analysis of the $B^0_d\to J/\psi \pi^0$ decay as a guideline,  
$a=0.41$ and $\theta=194^{\circ}$; for $\gamma$, we assume a value
of $65^\circ$. The resulting statistical errors are found as $\Delta C=\Delta S=0.14$ and $\Delta H= 0.069$ for 6 fb$^{-1}$  at a centre-of-mass energy of $\sqrt{s}=14$\,TeV (end of 2014--15), and 
$\Delta C=\Delta S=0.035$ and $\Delta H= 0.052$ for 100 fb$^{-1}$  (LHCb upgrade scenario).

\begin{figure}[t]
\centering
~\hfill
\includegraphics[width=0.34\textwidth]{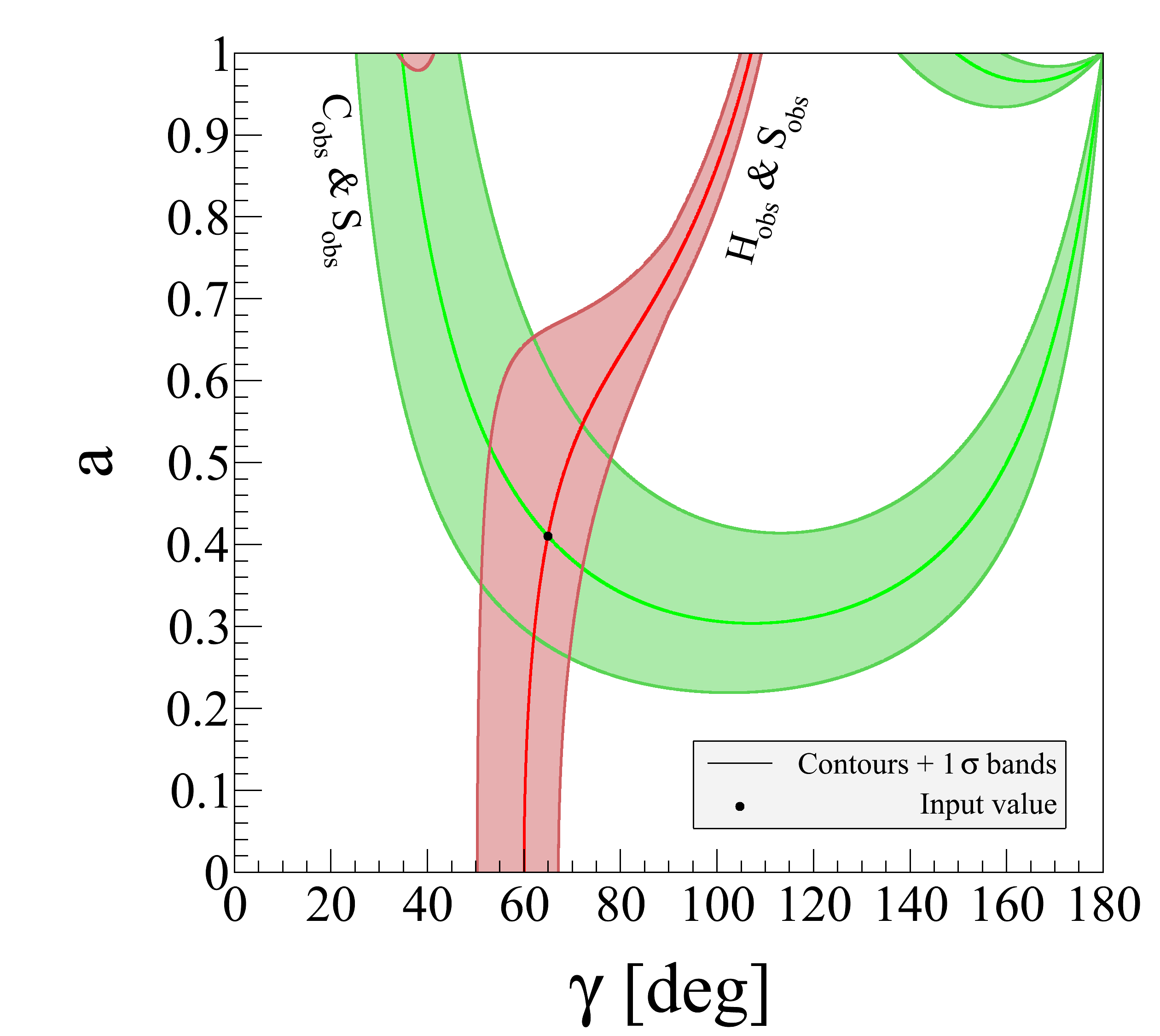}
\hfill
\includegraphics[width=0.34\textwidth]{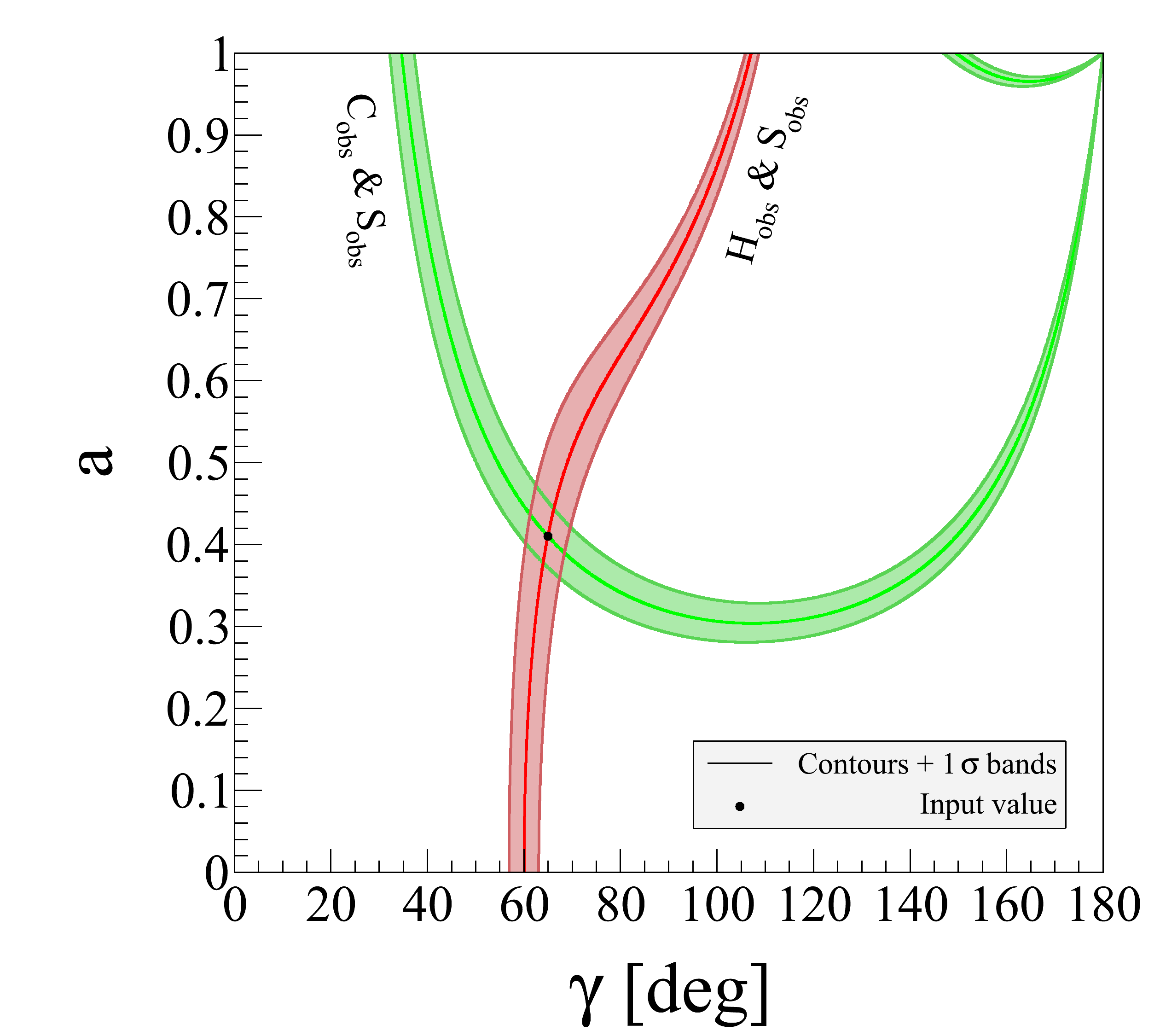} 
\hfill~
\caption{Determination of $\gamma$ and $a$ through intersecting contours, resulting 
from our LHCb feasibility study for 6 fb$^{-1}$ (left) and 100 fb$^{-1}$ (right).}
\label{fig:contours_gamma}
\end{figure}

Using the $B^0_s$--$\bar B^0_s$ mixing phase as an input, we can determine 
contours in the $\gamma$--$a$ plane, as illustrated in Fig.~\ref{fig:contours_gamma}. 
Their intersection allows us to pin down 
both $\gamma$ and the penguin parameters, and result in a statistical error for 
$\gamma$ of $10^{\circ}$ and $3.2^{\circ}$ for our two LHCb scenarios.

Although these precisions for $\gamma$ cannot compete with those from measurements
of pure tree decays at LHCb, which give errors about three times smaller, the important 
aspect of this $\gamma$ determination is that it may reveal a  New-Physics (NP) contribution 
to the $B^0_s\to J/\psi K_{\rm S}$ decay amplitude. In the following, we assume that we will
obtain a picture consistent with the Standrad Model for $\gamma$, and that NP will only manifest itself in the
$B^0_{d,s}\to J/\psi K_{\rm S}$ system through contributions to $B^0_{d,s}$--$\bar B^0_{d,s}$
mixing.

\boldmath
\section{Controlling Penguin Effects}\label{sec:pen-cont}
\unboldmath
%%%%%%%%%%%%%%%%%%%%%%%%%%%%%%%%%%%%%%%%%%%%%%%%%%%%%%%%%%%%%%%%%%%%%%%%%%%%

\begin{figure}[htb]
\centering
~\hfill
\includegraphics[width=0.33\textwidth]{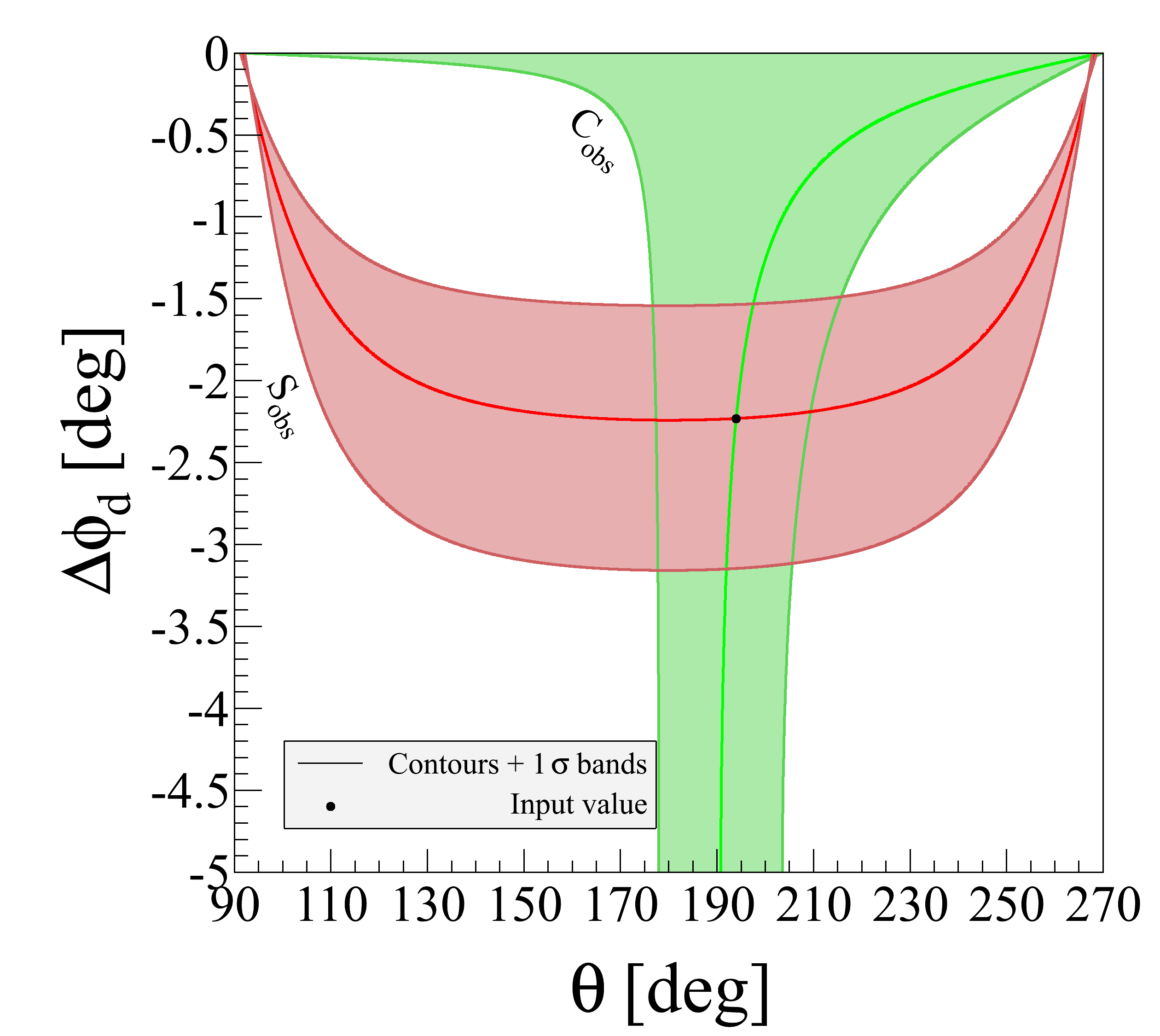}
\hfill
\includegraphics[width=0.33\textwidth]{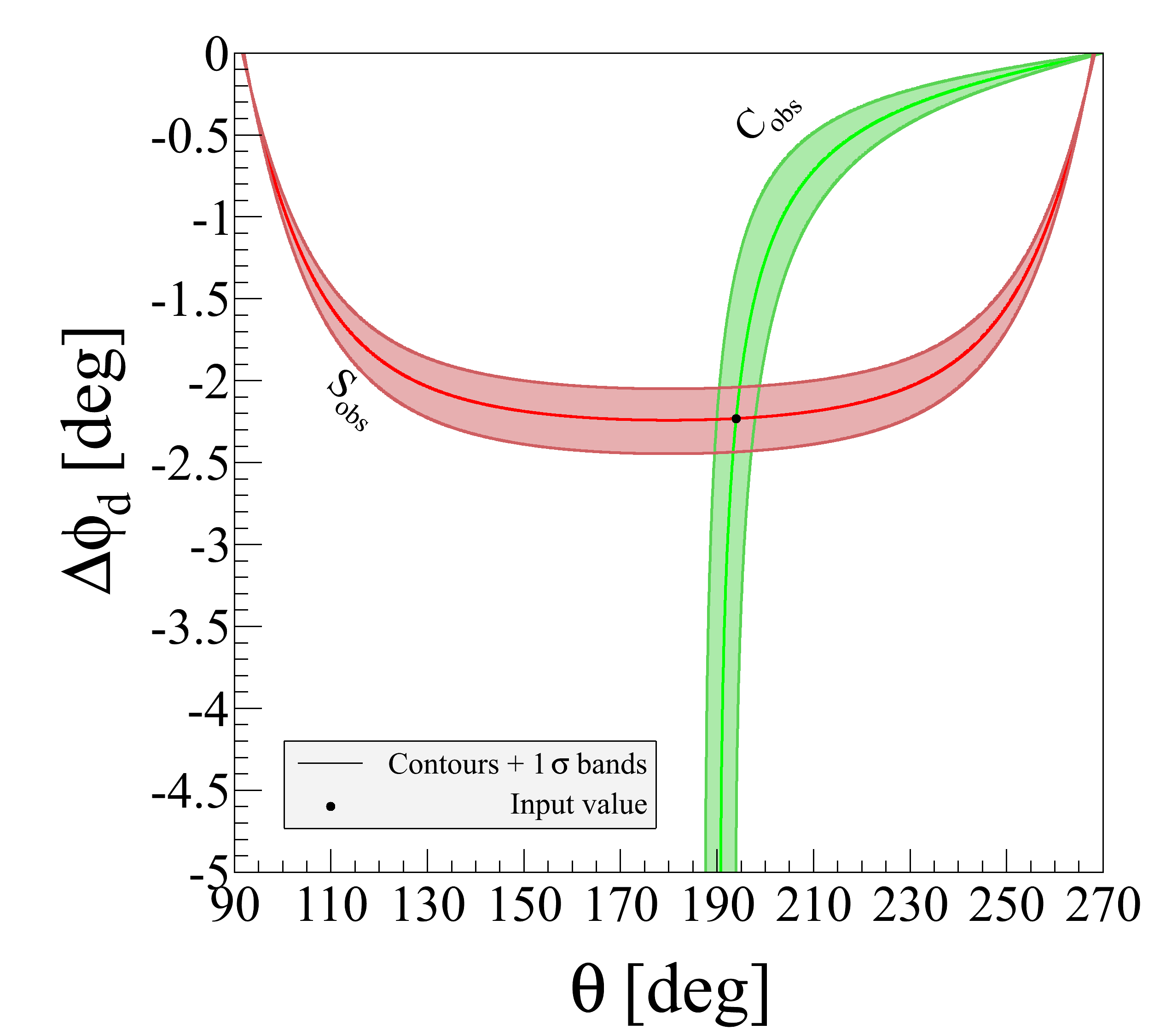} 
\hfill~
\caption{Determination of  $\Delta\phi_d$ (left panel: 6 fb$^{-1}$, right panel: 100 fb$^{-1}$).}
\label{fig:DelPhi}
\end{figure}

The major application of $B_s^0\to J/\psi K_{\rm S}$ at LHCb will be the extraction of the hadronic penguin parameters $(a,\theta)$ and their control in the determination of the 
$B^0_d$--$\bar B^0_d$ mixing phase $\phi_d$ from $B^0_d\to J/\psi K_{\rm S}$. 
The generalised expression for the measurement of this quantity reads \cite{FJFM}:
\begin{equation}
\frac{S(B_d^0\to J/\psi K_{\rm S})}{\sqrt{1-C(B_d^0\to J/\psi K_{\rm S})^2}}
=\sin(\phi_d+\Delta\phi_d),
\end{equation}
where the hadronic shift $\Delta\phi_d$ encapsulates the penguin topologies. Since our goal is to minimise the $U$-spin-breaking  corrections, we shall refrain from using the $H$ observable and will assume that $\gamma$ is known. In Fig.~\ref{fig:DelPhi}, we illustrate the determination of
$\Delta\phi_d$ for our specific example, yielding $\Delta\phi_d$ errors of 
$0.79^{\circ}$ and $0.19^{\circ}$ for  6 fb$^{-1}$ and 100 fb$^{-1}$, respectively \cite{Bs2JpsiKs}.

\begin{figure}[htb]
\centering
~\hfill
\includegraphics[width=0.33\textwidth]{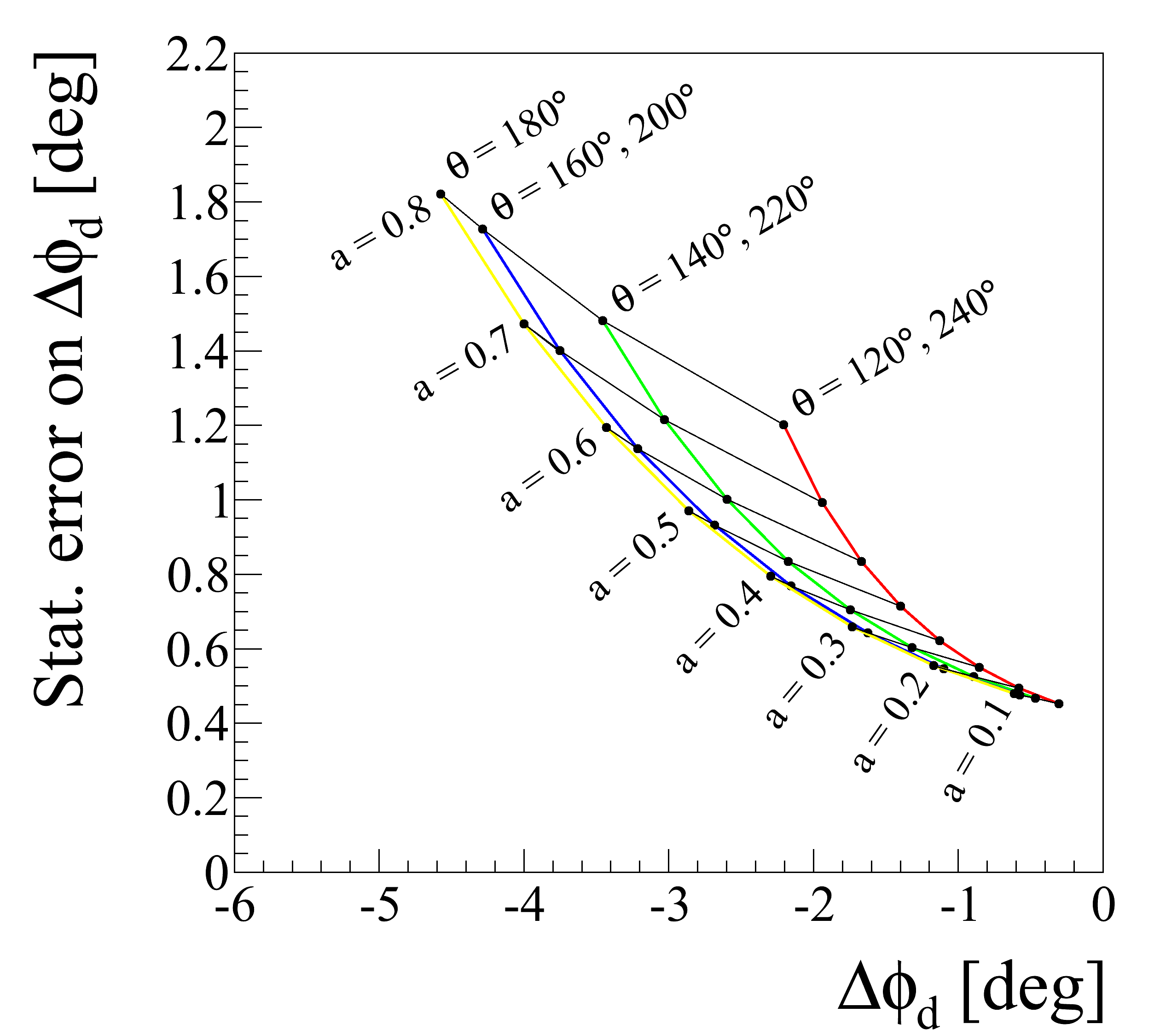}
\hfill
\includegraphics[width=0.33\textwidth]{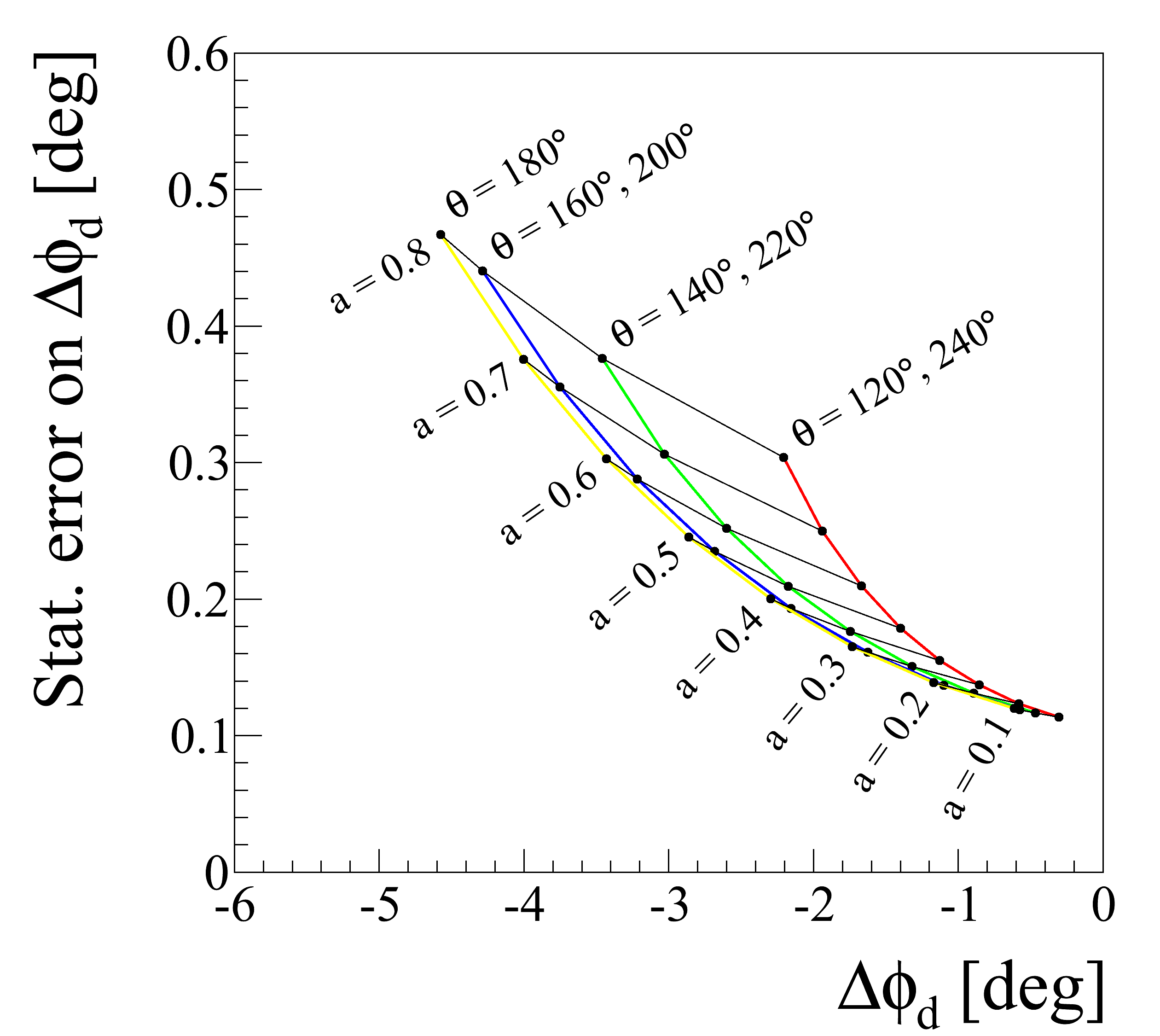} 
\hfill~
\caption{Correlation between $\Delta\phi_d$ and its error for our
LHCb feasibility study for different values of $a$ and $\theta$ (left: 6 fb$^{-1}$, 
right: 100 fb$^{-1}$).}\label{fig:DelPhi-corr}
\end{figure}

In order to study the dependence on the input values of the penguin parameters for our analysis, we show the correlation between the hadronic shift $\Delta\phi_d$ and its statistical error at LHCb in Fig.~\ref{fig:DelPhi-corr}. The corresponding curves show nicely that we can precisely determine the $\Delta\phi_d$ correction for a wide range of $(a,\theta)$ that should contain the ``true" values of these parameters. It should also be noted that already a small penguin contribution with $a=0.1$ gives a correction of $\Delta\phi_d\sim -0.5^\circ$.

In order to fully appreciate these results, $\Delta\phi_d$ should be compared with the expected statistical error on $\phi_d$ at LHCb, which is estimated as $1^\circ$ and $(0.2\mbox{--}0.8)^\circ$ for 6 fb$^{-1}$ and 100 fb$^{-1}$, respectively. In order to match these precisions, 
we definitely have to control the doubly Cabibbo-suppressed penguin contributions in 
$B^0_d\to J/\psi K_{\rm S}$. Looking at Fig.~\ref{fig:DelPhi-corr},  we observe that we can 
actually achieve this goal. Measurements along these lines may eventually allow us to 
resolve CP-violating NP contributions to $B^0_d$--$\bar B^0_d$ mixing.

\section{Physics with First LHCb Data}
\begin{figure}[htb]
\centering
~\hfill
\includegraphics[width=0.36\textwidth]{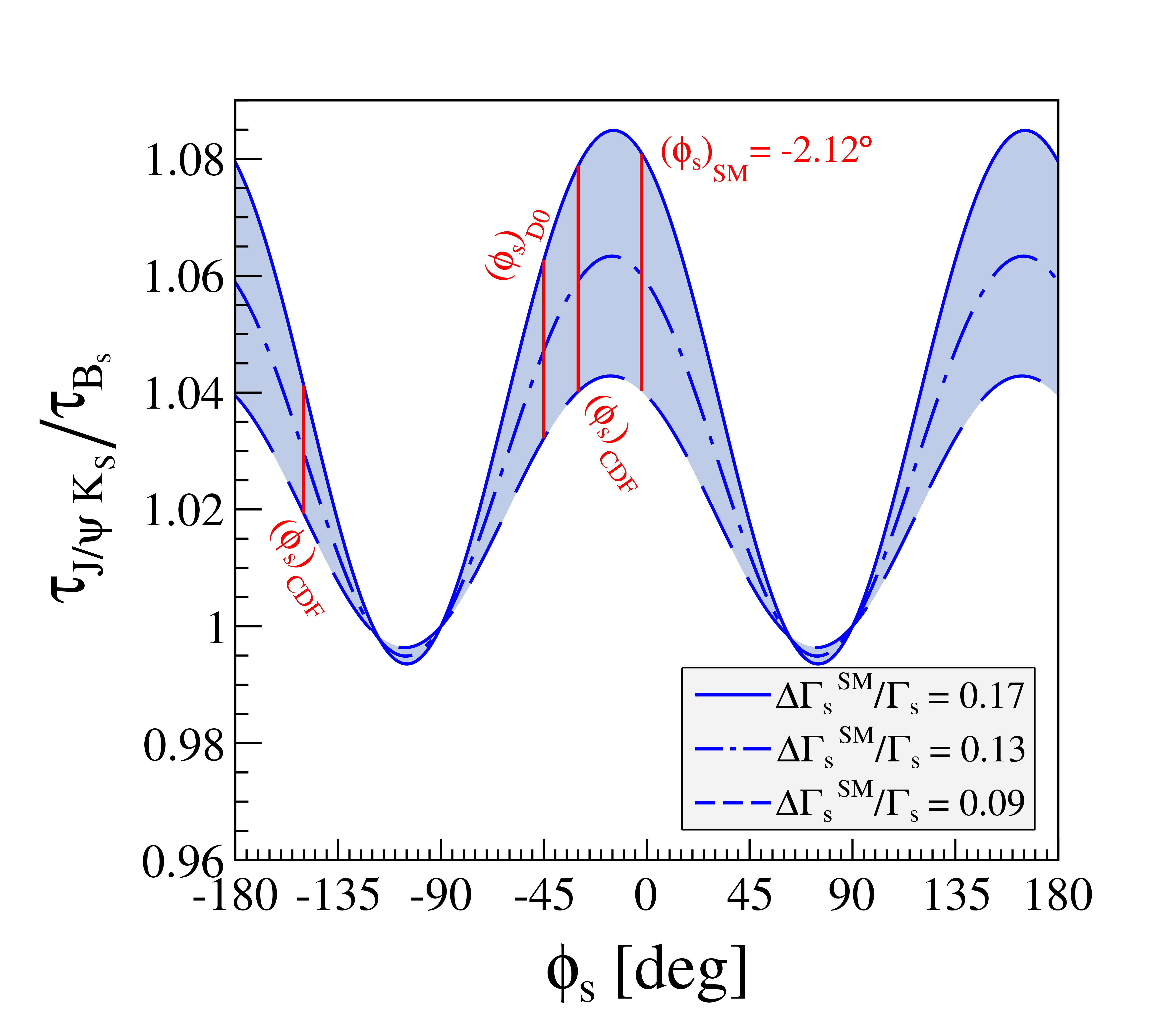}
\hfill
\includegraphics[width=0.36\textwidth]{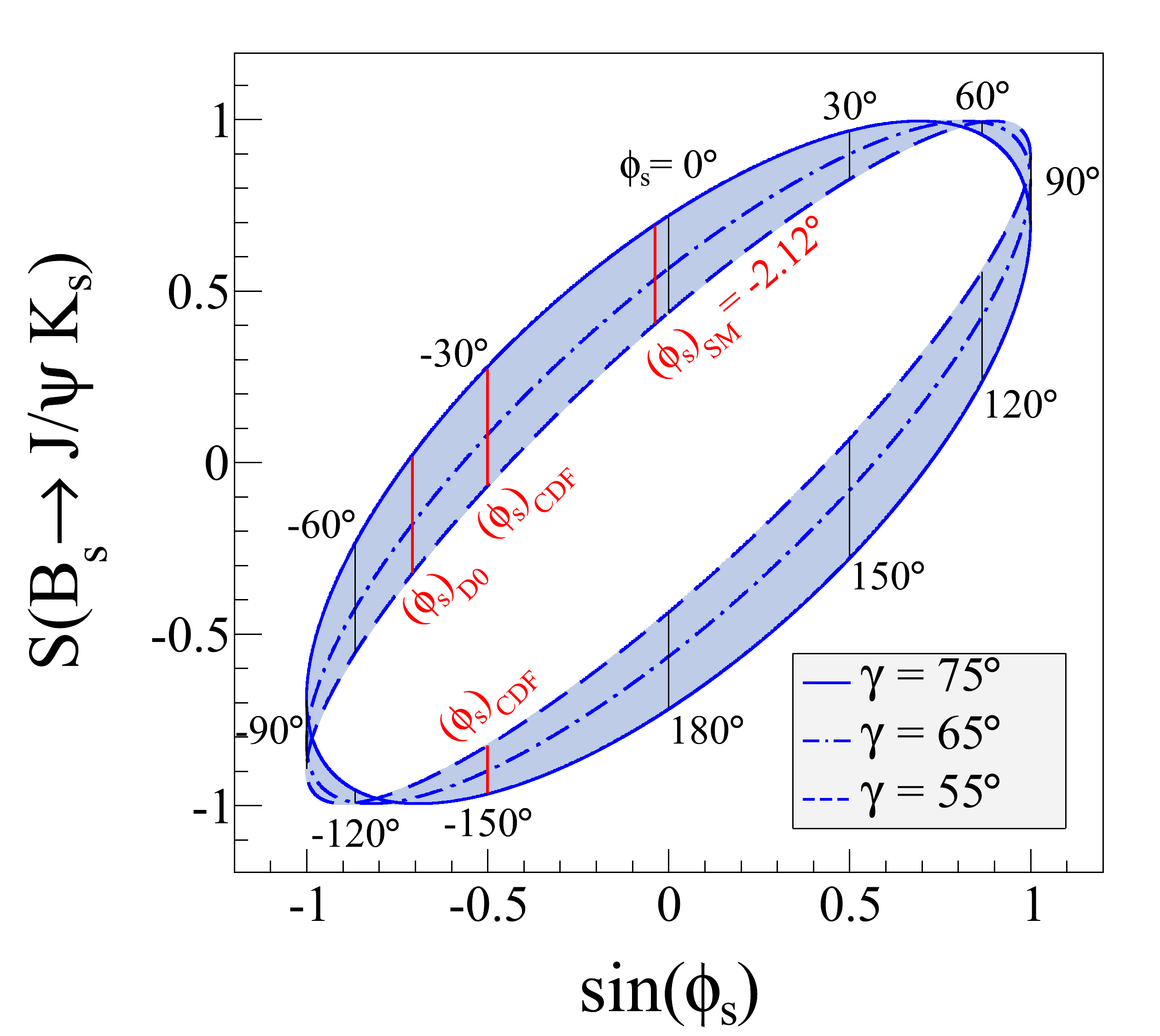}
\hfill~
\caption{Left panel: dependence of the effective $B^0_s\to J/\psi K_{\rm S}$ lifetime on $\phi_s$,
right panel: correlation between $S(B^0_s\to J/\psi K_{\rm S})$ and $\sin\phi_s$ for 
$\gamma=65^\circ$.}
\label{fig:corr}
\end{figure}

%%%%%%%%%%%%%%%%%%%%%%%%%%%%%%%%%%%%%%%%%%%%%%%%%%%%%%%%%%%%%%%%%%%%%%%%%%%%
An interesting target for first LHCb physics results on $B^0_s\to J/\psi K_{\rm S}$ is its effective
lifetime $\tau_{J/\psi K_{\rm S}}$. In the SM, we find 
\begin{equation}
\tau_{J/\psi K_{\rm S}}/\tau_{B_s}=1.060\pm0.020|_{\rm \Delta\Gamma_s^{\rm SM}/\Gamma_s}
\pm 0.010_{\rm Input},
\end{equation}
where the last error corresponds to $a\in[0.15,0.67]$, $\theta\in[174^{\circ},213^{\circ}]$ 
and $\gamma=(65\pm10)^{\circ}$ \cite{FJFM}. In the left panel of Fig.~\ref{fig:corr}, we show the
dependence on the $B^0_s$--$\bar B^0_s$ mixing phase. In the right panel of this figure, 
we show the correlation between the mixing-induced  $B^0_s\to J/\psi K_{\rm S}$ 
CP asymmetry and $\sin\phi_s$, which can be determined from the time-dependent 
angular analysis of the $B_s^0\to J/\psi\phi$ channel. For a similar analysis of 
$B^0_s\to K^+K^-$ and more details on the effective lifetime, the reader is referred to 
Ref.~\cite{FK}.

We look forward to confronting these results with first data at LHCb! By the end of 2011
(1fb$^{-1}$), a $B^0_s\to J/\psi K_{\rm S}$ signal should be clearly visible, with a sensitivity 
of $\Delta C=\Delta S=0.34$ for the CP-violating observables.

%%%%%%%%%%%%%%%%%%%%%%%%%%%%%%%%%%%%%%%%%%%%%%%%%%%%%%%%%%%%%%%%%%%%%%%%%%%%

\end{document}